\documentclass[twocolumn,preprintnumbers,amsmath,amssymb]{revtex4}

\usepackage{amsmath}
\usepackage{amssymb}
\usepackage{graphicx}% Include figure files

\begin{document}

\title{Quantum secured LiDAR with Gaussian modulated coherent states}% Force line breaks with \\
\author{Dong~Wang,$^{1,2}$ Juan-Ying~Zhao,$^{1,2}$ Ya-Chao~Wang,$^{1,2}$ Liang-Jiang~Zhou,$^{1,2,}$\footnote{ljzhou@mail.ie.ac.cn}
and Yi-Bo~Zhao$^{3,4,}$\footnote{zhaoyb@hizju.org}}
\address{$^1$ Aerospace Information Research Institute, Chinese Academy of Sciences, Beijing 100190, China\\
$^2$ National Key Laboratory of Microwave Imaging Technology, Beijing 100190, China\\
$^3$ Huzhou Institute of Zhejiang University, Huzhou 313000, China\\
$^4$ College of Control Science and Engineering, Zhejiang University, Hangzhou 310027, China}

\date{\today}% It is always \today, today,

\begin{abstract}
LiDAR systems that rely on classical signals are susceptible to intercept-and-recent spoofing attacks, where a target attempts to avoid detection. To address this vulnerability, we propose a quantum-secured LiDAR protocol that utilizes Gaussian modulated coherent states for both range determination and spoofing attack detection. By leveraging the Gaussian nature of the signals, the LiDAR system can accurately determine the range of the target through cross-correlation analysis. Additionally, by estimating the excess noise of the LiDAR system, the spoofing attack performed by the target can be detected, as it can introduce additional noise to the signals. We have developed a model for target detection and security check, and conducted numerical simulations to evaluate the Receiver Operating Characteristic (ROC) of the LiDAR system. The results indicate that an intercept-and-recent spoofing attack can be detected with a high probability at a low false-alarm rate. Furthermore, the proposed method can be implemented using currently available technology, highlighting its feasibility and practicality in real-world applications.
\end{abstract}
%\pacs{03.67.Dd, 03.67.Hk}
\maketitle

\section{Introduction}
Light detection and ranging (LiDAR) systems provide high-resolution and high-accuracy distance information of targets, thanks to their excellent angular and depth resolutions \cite{lidar1}. LiDAR has emerged as a crucial sensor in various applications, including reconnaissance, autonomous driving, and robotics. In most of these applications, a common concern and challenging issue is the threat of interference and background noise, where the receiver can be deceived by extraneous or false signals originated from the ambient light or intentional spoofing signals \cite{jam1,jam2,jam3}. To mitigate the impact of interference light and promote the detection security, several methods based on cross-correlation measurement were proposed, such as schemes based on a code-division-multiple-access (CDMA) technique \cite{cdma}, pseudo- or true-random bits modulations \cite{pseudo,true}, and laser chaos \cite{chaos}. Additionally, other countermeasures have been developed to protect LiDAR systems against spoofing attacks, which use side-channel information leaked from a cryptographic device \cite{AES} or frequency encrypted based on randomly varying chirp rates \cite{fencrypt}. Despite the excellent noise suppression capability \cite{anti-inter}, they remain vulnerable to spoofing attacks using an intercept-resend (IR) jamming strategy \cite{intercept}, on account of the fact that the classical signals emitted by LiDAR transmitter can be perfectly replicated in theory.

Benefit from the second quantum revolution \cite{2ndquantum}, the fields of targets detection and imaging have made significant progress with the help of quantum technologies. In these applications, entangled photons are commonly utilized to enhance object detection performance in the presence of spurious light and noise. In quantum illumination \cite{qi1,qi2} and some quantum secured imaging protocols \cite{qsi1,qsi2}, each signal sent out to irradiate the object is entangled with a retained idler. The non-classical correlation between the returning signal and its entangled pair is then measured, thereby improving the signal-to-noise and the resistance to jamming \cite{qi3,qi4,qsi3,qsi4}. However, the efficient generation and detection of entangled light, which are essential for practical systems, still pose a challenge with current technology.

As an alternative, quantum secured imaging and LiDAR protocols based on single photons have been proposed \cite{qsl1,qsi1,qsl2} to ensure detection security. In these protocols, photons that query a target are randomly encoded into four BB84 polarization states. The echoes from the target are then detected by single-photon detectors. Any attempt to jam the system by intercepting and resending the photons with false information will introduce statistical errors. This is due to the fact that a photon cannot be measured simultaneously in two conjugate polarization bases \cite{qsi1}. To meet the security requirements, the transmitter is expected to emit a few photons, ensuring that less than one photon per pulse reaches the target. In more conservative cases, pulses at single photon level are emitted directly. However, in real-life implementations, attenuated lasers containing multi-photon components are often used, as demonstrated in proof-of-principle experiments \cite{qsi1,qsl2,qsi5}. In consequence, the security of the system can be compromised.

To circumvent the defects of the single-photon regime, we turn to quantum continuous variables of coherent states, which also play an important role in high-rate and cost-effective quantum key distribution \cite{cv1,cv2,cv3}. Limited by the Heisenberg uncertainty principle, it is impossible to precisely measure both canonically conjugate quadratures of a coherent state simultaneously. This means that an IR spoofing attack will always introduce extra noise to a LiDAR system that uses coherent states, making it possible to detect the attack. Based on this fundamental prerequisite, we propose a practical Gaussian modulation quantum secured LiDAR (GMQSL) protocol. In our protocol, the LiDAR transmitter modulates the position quadrature ($X$) and momentum quadrature ($P$) of a coherent state in phase space using a bivariate Gaussian distribution. The modulated state is then sent to a non-cooperative target. The receiver randomly measures one of the two quadratures using homodyne detection. The cross-correlation of the transmitted and received Gaussian data is used for ranging, and the system's noise is estimated for security check. 

We've noticed that a similar concept was explored preliminary in ref. \cite{qlimit}. However, the impact of loss and noise was neglected, and no specific implementation was discussed.  In contrast, our current work proposes a practical quantum secured LiDAR system based on Gaussian-modulated coherent states. The hardware required for this system is almost identical to that of a traditional coherent LiDAR and can be realized using readily available components. Furthermore, we also analyze the security and performance of our system in the presence of loss and noise, particularly under an IR spoofing attack. Simulation results demonstrate the feasibility and practicality of our approach.

This paper is organized as follows. In Sec. \ref{sec2}, we introduce the proposed GMQSL protocol. In Sec. \ref{sec3}, we describe the signal model of the LiDAR system. In Sec. \ref{sec4}, we present the operating principle for target detection and security check of the system. In Sec. \ref{sec5}, we analyze the performance of the system through numerical simulations. Lastly, in Sec. \ref{sec6}, we give concluding remarks.

\section{\label{sec2}Protocol description}
Practically speaking, existing spoofing technologies in target detection typically employ a “measure and prepare” strategy. In this strategy, the target or spoofer intercepts and measures the signals transmitted by the LiDAR transmitter, then faithfully reproduces these signals but with false information based on their measurement results, which are sent to the LiDAR receiver. Fortunately, quantum mechanics imposes a fundamental limit on the information that can be obtained by the spoofer through a single measurement on the transmitted signal. This means that if the LiDAR transmitter randomly chooses non-orthogonal quantum states, the spoofer can only determine the quantum state by using classical information from the measurement. 

Since a coherent state $\left|\alpha\right\rangle$ has the minimum uncertainty, just as a vacuum state \cite{qoptics}, a spoofer cannot measure the exact value of its quadrature $X$ and quadrature $P$ simultaneously. Consequently, the spoofer fails to generate a spoof signal that closely resembles the transmitted signal. In fact, if the coherent states are distributed in phase space according to a Gaussian distribution, the optimal fidelity of the reconstructed states through a measure-and-prepare strategy is $1/2$, which can be achieved by heterodyne detection \cite{benchmark}. To make the spoof as convincing as possible, we assume that the spoofer performs a heterodyne measurement of the transmitted coherent state to carry out the IR spoofing attack. As a result, this introduces extra noise to the system that is at least twice the shot-noise unit (SNU) \cite{clone}, which is significant compared to other sources of noise.

Interestingly, Gaussian distributed signals are well-suited for ranging purposes due to their excellent atuo-correlation properties. Additionally, since coherent states can be easily generated using a laser, and simple homodyne detection has been proven to be near-optimal among all available Gaussian measurements \cite{homo}, we can straightforwardly design a practical quantum-secured LiDAR protocol called the GMQSL protocol.

\begin{figure*}[!ht]
\centering
\includegraphics[width=0.8\linewidth]{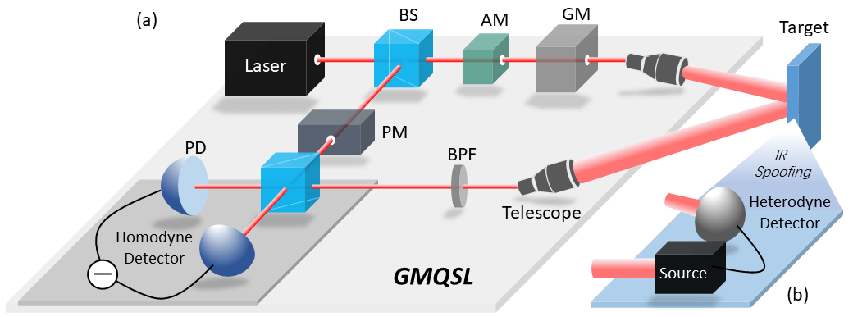}
\caption{a. A schematic diagram of GMQSL; b. IR spoofing attack. BS, beam splitter; AM, amplitude modulator; PM, phase modulator; GM, Gaussian modulator; PD, photondiode; BPF, band-pass filter. Homodyne detector consists of a BS and two PDs; Heterodyne detector consists of a BS and two homodyne detectors.}
\label{scheme}
\end{figure*}

An example scenario for performing the GMQSL protocol is depicted in Fig.\ref{scheme}. In this scenario, a continuous wave laser is split into two parts using a beam splitter (BS). One part is used to generate amplitude-modulated signal pulses through an amplitude modulator, while the other part serves as the local oscillator (LO) in the homodyne detection. The signal pulses are further modulated by a Gaussian modulator with Gaussian variables. This modulation can be achieved by cascading an amplitude modulator with a Rayleigh distribution and a phase modulator with a uniform distribution, resulting in zero-centered Gaussian distributions. 

The quantum signals are then transmitted to a target through a telescope. The target has the capability to perform an IR spoofing attack by simultaneously measuring both quadratures of the signals using a heterodyne detector. It can then generate fake signals using a light source and send them to the receiver. The returned signals are collected by another telescope and filtered using a band-pass filter (BPF). These signals are then interfered with the LO on a shot-noise-limited balanced homodyne detector, which consists of a second beam splitter and two photodiodes. The LO path includes a phase modulator, allowing for the random choice of the measured signal quadrature.

The general principle of the GMQSL protocol, along with a detailed procedure, can be summarized as follows:

(1)Random number generation: The LiDAR system generates two sets of independent Gaussian random variables, $X_T$ and $P_T$, with a centered bivariate Gaussian modulation of variance $V_M$. Each set has a length of $M$. Additionally, a series of binary random variables, $B$ is generated with a length of $M'$, where $M'>M$.

(2)Quantum signals preparation and target irradiation: The transmitter prepares $M$ quantum states, $\left| \alpha \right\rangle=\left|{{X_T}+i{P_T}}\right\rangle$, by modulating $X_T$ and $P_T$ to the quadrature $X$ and quadrature $P$ of coherent states generated by a laser. These states are then sent to a target.

(3)Echoes detection and data acquisition: The receiver collects the echoes from the target and randomly selects a bit from $B$ to determine whether to measure the $X$ or $P$ quadrature. This is done by performing a balanced homodyne detection on the selected quadrature. The measurement results, denoted as $R$, are recorded as $X_R$ or $P_R$ depending on the bit value of $B$.

(4)Ranging and data processing: A subset of $L$ states is chosen for comparison with the measured results by calculating the moving cross-correlation. For each bit, two new sequences, $T_1 = (1-B) \cdot X_T + B \cdot P_T$ and $T_2 = B \cdot X_T - (1-B) \cdot P_T$, are constructed based on $X_T$, $P_T$ and $B$. The cross-correlations between $R$ and $T_1$, $T_2$ are calculated, respectively, and the maximum absolute values is used for ranging. The total sent and measured sequences are then aligned and followed by phase compensation.  

(5)Security check: The excess noise of the system is estimated using the processed data and compared with a predetermined threshold to check for security. If the excess noise exceeds the threshold, it indicates a potential spoofing attack.

Note that in real implementations, apart from the inherent shot-noise in the receiver's result, there are various sources of noise such as imperfect state preparation, transmission, background light and detection. The last source is referred to as detection noise, which is assumed to be uncontrollable by the spoofer and can be characterized beforehand. The other sources of noise are defined as excess noise referred to the transmitter output and may change due to the IR spoofing attack. By evaluating the excess noise, it is possible to determine whether a spoofing attack has taken place.

\section{\label{sec3} Signal model}

The prepared coherent states can be expressed in terms of phase $\theta$ and amplitude $A$. The distribution of the phase corresponds to a uniform modulation in the range of $[0,2\pi]$, while the amplitude follows a Rayleigh distribution with a variance of $V_M$. The quadrature $X$ and quadrature $P$ can be written as
\begin{equation}\label{XT}
X_T=A\cos\theta,
P_T=A\sin\theta.
\end{equation}
with $X_T,P_T \sim N(0,V_M)$ and $V_M=\left\langle {{A^2}{{\cos }^2}\theta }\right\rangle =\left\langle {{A^2}{{\sin}^2}\theta }\right\rangle $, where $\left\langle \cdot \right\rangle $ represents the expectation.

The quantum signals then undergo a round trip propagation between the LiDAR system and the target through a free-space channel.
%Since the channel fluctuations are on the order of kHz, while the signal repetition rate can reach tens of MHz or even higher, at least tens of thousands of probe states can be transmitted during the stability time of the fading atmospheric channel \cite{fading}. 
\subsection{Absence of spoofing attack}
In the absence of a spoofing attack, the transfer of the $X$-quadrature can be described in the Heisenberg picture by the following evolution (the same applies to the $P$-quadrature):
\begin{equation}\label{XR1}
\begin{aligned}
{X_{Ra}} &= \sqrt {{\eta _d}} \left[\sqrt {\eta _c^2{\eta _r}} \left( {{X_T} + {X_{0}}} + {X_{\varepsilon} }  \right) + \sqrt {1 - \eta _c^2{\eta _r}} {X_{1}} \right] \\
&+ \sqrt {1 - {\eta _d}} {X_{2}}  + {X_d},
\end{aligned}
\end{equation}
where all variables are normally distributed with zero mean, except for the detector efficiency $\eta_d$, channel transmittance $\eta_c$ and target reflectivity $\eta_r$. $X_{Ra}$ represents the measured quadrature of the state; $X_T$ denotes the value of the Gaussian displacement of the signal with modulation $V_M$; $X_{0}$, $X_{1}$ and $X_{2}$ represents the vacuum noise with unity variance (also known as shot-noise unit); $X_{\varepsilon}$ ($X_d$) corresponds to the excess (detector) noise with a variance of $\xi_0$ ($V_d$).

It should be noted that in practice, $X_{0}$, $X_{1}$, $X_{2}$, $X_\varepsilon$, and $X_d$ are all independent Gaussian noises (and are unknown quantities). The variance of $X_{Ra}$ can be calculates as
\begin{equation}\label{VR}
\begin{aligned}
{V_{Ra}} &=\eta_a\left( {{V_M} + 1 + {\xi_0}} \right) + \left( {{\eta _d} - {t_a^2}} \right) + \left( {1 - {\eta _d}} \right) + {V_d} \\
& =\eta_a{V_M} + V_{Na},
\end{aligned}
\end{equation}
where $\eta_a={{\eta_d}\eta _c^2{\eta _r}}$ represents the total channel transmission efficiency, and $V_{Na}=\eta_a{\xi_0} + {V_d} + 1$ is the total noise variance. All variances are normalized to shot-noise unit. The signal-to-noise ratio (SNR) of the measured results can be expressed as
\begin{equation}\label{SNRa}
\text{SNR}_a=\frac{\eta_aV_M}{V_{Na}}.
\end{equation}

\subsection{Presence of spoofing attack}

To conduct the IR spoofing attack, the spoofer is assumed to measure the quadratures of the coherent states transmitted by the LiDAR system using a heterodyne detector. The spoofer then prepares new coherent states based on its measurement results and sends them to the receiver. In this scenario, the spoofer utilizes a perfect heterodyne detector with unity detection efficiency and no electronic noise. The quadratures $X$ and $P$ are measured simultaneously by the spoofer, and the results are expressed as
\begin{equation}
\begin{aligned}
X_s & = \sqrt{\frac{\eta_c}{2}} \left(X_T + X_{0} + X_{\varepsilon} \right) + \sqrt{1-\frac{\eta_c}{2}} X_{3},\\
P_s & = \sqrt{\frac{\eta_c}{2}} \left(P_T + P_{0} + P_{\varepsilon} \right) + \sqrt{1-\frac{\eta_c}{2}} P_{3},
\end{aligned}
\end{equation}
where $X_{0}$ ($P_{0}$) represents the noise-term due to coherent-state modulation, while $X_{3}$ ($P_{3}$) represents the noise-term due to a 3-dB loss in the heterodyne detection and the channel loss from the transmitter to the spoofer. $X_{\varepsilon}$ ($P_{\varepsilon}$) corresponds to the excess noise with a variance of $\xi_0$. 

In the repreparation stage, the spoofer prepares a coherent state with quadratures $X_p$ and $P_p$ based on its measurements $X_s$ and $P_s$. The spoofer can also induce an amplification or attenuation of the data $X_s$ with a gain of $\gamma$ before optical modulation. In our further analysis, we will fous on the quadrature $X$, but the treatment for the quadrature $P$ is totally symmetric. The resent quadrature can be written as
\begin{equation}
X_p = \sqrt{\gamma}X_s + X_{4},
\end{equation}
where $X_{4}$ is the noise term due to the coherent-state preparation.

The coherent states reprepared by the spoofer eventually reach the receiver and are measured using homodyne detection, the measured result $X_{Rp}$ can be written as
\begin{equation}
\begin{aligned}
X_{Rp} & =\sqrt {{\eta _c}{\eta _d}} \left( {{X_p} + \sqrt {\frac{{1 - {\eta _c}}}{{{\eta _c}}}} {X_{5}}} \right) \\
& + \sqrt {1 - {\eta _d}} {X_{6}} + {X_d}  \\
&=\sqrt{\eta_p} X_T + X_{Np},
\end{aligned}
\end{equation}
where ${\eta_p}={{{\gamma\eta _c^2{\eta _d}} \mathord{\left/  {\vphantom {{k\eta _c^2{\eta _d}} 2}} \right.  \kern-\nulldelimiterspace} 2}} $ represents the effective channel transmission efficiency for the reprepared coherent states. $X_{Np}$ represents the total noise including $X_{0}$, $X_{3}$, $X_{4}$, $X_{5}$, and $X_{6}$ with unity variance each, as well as the excess noise $X_{\varepsilon}$ and detector noise $V_d$. The variance of $X_{Rp}$ can be written as
\begin{equation}\label{VRs}
\begin{aligned}
{V_{Rp}} &= {\eta_p}\left( {{V_M} + {\xi_0 + \frac{2}{{{\eta _c}}}}} \right) + {V_d} + 1 \\
& = {\eta_p}{V_M} + V_{Np},
\end{aligned}
\end{equation}
where $V_{Np}={\eta_p} \xi_1 + V_d + 1$, and $\xi_1=\xi_0+\frac{2}{{{\eta _c}}}$. The SNR is given by
\begin{equation}\label{SNRp}
\text{SNR}_p=\frac{{\eta_p}V_M}{V_{Np}}.
\end{equation}

We can now formalize a general signal model for the LiDAR system, considering both the absence and presence of a spoofing attack. The signal model is given by
\begin{equation}\label{sigmodel}
\begin{aligned}
X_R&=\sqrt{\eta} X_T+X_N,\\
V_N&=\eta \xi + V_d + 1
\end{aligned}
\end{equation}
where $X_R$ represents the measured quadrature at the receiver, $X_T$ is the transmitted quadrature, and $X_N$ is the total noise. In the absence of a spoofing attack, the total channel transmission efficiency is denoted by $\eta_a$ and the excess noise is denoted by $\xi_0$. In the presence of a spoofing attack, the total channel transmission efficiency is denoted by $\eta_p$ and the excess noise is denoted by $\xi_1$. The total noise $X_N$ follows a centered normal distribution with a variance determined from the observed data, denoted by $V_N$. We assume perfect phase compensation in this model. 

\section{\label{sec4} Operating principle for target detection and security check}
\subsection{Target detection}

Taking into account the influence of phase drift during atmospheric channel propagation, we can simplify Eq.(\ref{XR1}) as follows:
\begin{equation}\label{XR2}
X'_R =tA\cos\left(\theta+\delta  \right) + X_N,
\end{equation}
where $\delta $ represents the phase drift. Then the cross-correlation between the transmitted and measured sequences can then be calculated as
\begin{equation}\label{cc1}
Cov\left({X_T}X{'_R}\right) =\left\langle {{X_T}X{'_R}} \right\rangle =\sqrt{\eta}V_M\cos\delta  ,
\end{equation}
Where $Cov\left(\cdot\right)$ represents the covariance. Analogously, for $P$-quadrature we have $Cov\left({P_T}P{'_R}\right)=\sqrt{\eta}V_M\cos\delta $. It is evident that the cross-correlation value obtained from each quadrature can be greatly affected by the phase drift. For example, the cross-correlation value of the $X$-quadrature reaches its maximum when $\delta =0$, while drops to 0 when $\delta =\pi/2$. This makes ranging a formidable task, as the cross-correlation value is highly sensitive to the phase drift.

To tackle this problem, we take advantage of the correlations bwtween the transmitted and measured sequences from conjugate quadratures. These correlations can be easily obtained as follows
\begin{equation}\label{cc2}
\begin{aligned}
Cov\left({X_T}P{'_R}\right) & = \sqrt{\eta}V_M\sin\delta , \\
Cov\left(-{P_T}X{'_R}\right) & = \sqrt{\eta}V_M\sin\delta .
\end{aligned}
\end{equation}

By constructing two new sequences $T_1 = (1-B) \cdot X_T + B \cdot P_T$ and $T_2 = B \cdot X_T - (1-B) \cdot P_T$ based on $X_T$, $P_T$ and $B$, we can calculate the peak value of the cross-correlation of $R$ and $T_1$ ($T_2$)  (see Appendix \ref{app1})
\begin{equation}\label{cc3}
\begin{aligned}
C^p_{{}T_\nu}{R}:&= Cov\left({T_\nu}R\right)  = \sqrt{\eta}{V_M}\mu_\nu, \\
{\mu _\nu} &= \left\{ {\begin{array}{*{20}{c}}
{\cos \delta , \text{if } \nu = 1}\\
{\sin \delta , \text{if } \nu = 2}
\end{array}} \right..
\end{aligned}
\end{equation}

To ensure robust ranging in the presence of phase drift, we compare the absolute values of the cross-correlations peak and select the larger one, denoted as ${C_{\max }} =\sqrt{\eta}V_M \mathop {\max }\limits_{\nu = 1,2} \left\{ {\left| {{\mu _\nu}} \right|} \right\}$. Therefore, the normalized peak value of the cross-correlation remains no less than $1/\sqrt{2}$ regardless of the phase drift. 

As a result, ranging can be achieved by performing cross-correlation operations between the constructed sequence $T_1$ (or $T_2$) and the received sequence $R$, shifting bit by bit. By selecting the cross-correlation with the larger absolute value, the system can effectively handle arbitrary phase drift.

However, in a real-life implementation, we need to consider the finite-size effect due to the finite length $L$ of the sequences. In this case, the cross-correlation outcome will fluctuate due to finite measurements. Denote the realizations of $T_1$, $T_2$ and $R$ with $T_{1i}$, $T_{2i}$ and $R_i$ $(i \in \{1,2, \cdots,L \})$, respectively. The covariance between $T_\nu$ and $R$ can be estimated using the maximum-likelihood estimator
\begin{equation}
\widehat {C_{{}T_\nu}{R}} = \frac{1}{L}\sum\limits_{i = 1}^L {{T_{\nu i}}{R_i}},\nu=1,2.
\end{equation}

When the transmitted and received sequences are misaligned, they become completely independent and act as the noise floor of the cross-correlation profile. In this case, the cross-correlation follows a centered normal distribution, given by (see Appendix \ref{app2}):
\begin{equation}\label{Cnf}
\widehat {{C^{nf}_{{T_\nu}R}}} \sim \mathcal N \left(0, V^{nf}_C \right),
\end{equation}
where 
\begin{equation}
V^{nf}_C=\frac{{{\eta}V_M^2}}{L}\left( {1 + \frac{1}{\text{SNR}}} \right)
\end{equation}

On the other hand, when the transmitted and received sequences are aligned, the cross-correlation reaches a peak, whose estimator obeys the following distribution (see Appendix \ref{app3}):
\begin{equation}\label{Cp}
\widehat {{C^p_{{T_\nu}R}}}  \sim \mathcal N\left( {\sqrt{\eta}V_M\mu_\nu ,V^p_{C\nu} } \right),
\end{equation}
where
\begin{equation}
V^p_{C\nu}=\frac{{{\eta}V_M^2}}{L} \mu^2_\nu +V^{nf}_C.
\end{equation}

%Therefore, the performance of the system can be evaluated by the SNR and the target detection statistics. The SNR of the correlation profile can be denoted by correlation signal-to-noise ratio (CSNR) \cite{snr}, which is defined as the ratio of the power in the peak correlation to the variance in the noise floor, and is expressed as
%\begin{equation}\label{CSNR}
%\text{CSNR} = \frac{{{C_{\max %}^2}}}{{{V^{nf}_C}}}=\frac{L\mathop {\max %}\limits_{\nu = 1,2} \left\{ {\mu^2 _\nu} %\right\}}{1+1/\text{SNR}}.
%\end{equation}

%The gain of correlative signal processing can be defined as \cite{gain}
%\begin{equation}\label{GC}
%G_C=\frac{\text{CSNR}}{\text{SNR}}=\frac{L\mathop {\max }\limits_{\nu = 1,2} \left\{ {\mu^2 _\nu} \right\}}{1+\text{SNR}},
%\end{equation}
%which can be used to evaluate the SNR enhancement of the correlative processing of sent and received signals.

The statistics of target detection in the absence of a spoofing attack can be characterized by the probability of false alarm, denoted as $P^T_\text{fa}$, and the probability of detection, denoted as $P^T_\text{d}$, using a correlation value threshold $C_{th}$. The probability of false alarm represents the likelihood that the system will report a measurement above $C_{th}$, which is due to the noise floor. On the other hand, the probability of detection describes the likelihood that the measurement peak is both the peak value of the correlation profile and above the threshold $C_{th}$. Referring to Eq. (\ref{Cnf}) and Eq. (\ref{Cp}), we have
\begin{equation}\label{TROC}
\begin{aligned}
P^T_\text{fa}& = \frac{1}{2} \text{erfc}\left( {\frac{{{C_{th}}}}{{\sqrt {2{V^{nf}_C}} }}} \right)  ,\\
P_\text{d}^T & = \frac{1}{2}\text{erfc}\left( {\frac{ {C_{th}} - {C_\text{max}} }{{\sqrt {2{V^p_{C\nu}}} }}} \right),
\end{aligned}
\end{equation}
where 
\begin{equation}
\text{erfc}(x)=\frac{2}{{\sqrt \pi  }}\int_x^\infty  {{e^{ - {z^2}}}dz}
\end{equation}
is the complementary error function. 

After completing the ranging step, the transmitted and received sequences are aligned. To estimate the excess noise for security checks, the data needs to undergo further processing with phase compensation.

Since Eq. (\ref{cc3}) includes the phase drift parameter, we can obtain its tangent by simply calculating
\begin{equation}\label{ps}
\tan \delta  = \frac{{C_{{T_2}R}}}{{C_ {{T_1}R}}}.
\end{equation}
By considering the sign of the correlation value in Eq.(\ref{cc3}), we can determine the value of the phase drift. With this information, we can retrieve the linear relationship indicated in Eq. (\ref{sigmodel}) by rotating the phase of the data to $\theta +\delta $.

\subsection{Security check}
In this section, we analyze the security model of the GMQSL protocol under the IR spoofing attack.

By comparing the expressions of $V_{Na}$ and $V_{Np}$, we can clearly see that the IR spoofing attack will introduce an excess noise of ${2 \mathord{\left/  {\vphantom {2 {{\eta _c}}}} \right.  \kern-\nulldelimiterspace} {{\eta _c}}}$ shot-noise units, which depends solely on the channel transmittance $\eta _c$ only. As a result, the system is able to detect the spoofer by estimating the excess noise $\xi$ and comparing it with a threshold value $\xi_{th}$. 

Next, we will describe how to detect spoofing through parameters estimation. 

It is more straightly to estimate $V_\varepsilon={\eta}\xi$, as it can be conveniently estimated using the noise variance, where the term $V_d+1$ can be accurately characterized beforehand. Therefore, it is necessary to estimate ${\eta}$ in order to obtain the threshold value $V^{th}_\varepsilon={\eta}\xi_{th}$.

In the finite-size scenario, we will now describe the method for parameter estimation using $M$ interrogating signals to estimate $V_\varepsilon$. Denote the realizations of $T_1$ and $R$ with $T_{1i}$ and $R_i$, respectively. The maximum-likelihood estimators for $\sqrt{\eta}$ and $V_N$ are given by
\begin{equation}
\widehat {\sqrt{\eta}} = \frac{{\sum\nolimits_i^M {{T_{1i}}{R_i}} }}{{\sum\nolimits_i^M {T_{1i}^2}}}, 
\widehat {{V_N}}  = \frac{1}{M}\sum\limits_{i = 1}^M {{{\left( {{R_i} - \widehat {\sqrt{\eta}}{T_{1i}}} \right)}^2}} ,
\end{equation}
which comply with the normal distribution and chi-square distribution \cite{finite1}, respectively,
\begin{equation}
\widehat {\sqrt{\eta}} \sim N\left( {{\sqrt{\eta}}, \frac{{{V^p_{C1}}}}{{V_M^2}} } \right),\frac{{M\widehat {{V_N}}}}{{{V_N}}} \sim {\chi ^2}\left( {M - 1} \right),
\end{equation}
and the variance of estimator $\widehat {V_N}$ is given by
\begin{equation}
D\left( {\widehat {{V_N }}} \right) = \frac{{2V_N^2}}{M},
\end{equation}
where $D(\cdot)$ represents the variance.

Then the maximum-likelihood estimator for $V_\varepsilon$ and the corresponding variance can be written as \cite{finite2}
\begin{equation}
\widehat {V_\varepsilon} = \widehat {{V_N}} - (1+V_d), D\left( {\widehat {{V_\varepsilon }}} \right) = \frac{{2V_N^2}}{M}.
\end{equation}
Now we can estimate $V_\varepsilon$ for both case of spoofing absence or presence, which are denoted as $V^{a}_\varepsilon$ and $V^p_\varepsilon$, respectively.

Analogous to the case of target detection, we use the probability of false alarm and detection for security check, 
\begin{equation}
\begin{aligned}
P^S_\text{fa}& = \frac{1}{2} \text{erfc}\left( {\frac{{{V^{th}_\varepsilon-V^{a}_\varepsilon}}}{{\sqrt {2{D(V^{a}_\varepsilon)}} }}} \right)  ,\\
P_\text{d}^S & = \frac{1}{2}\text{erfc}\left( {\frac{{ {V^{th}_\varepsilon} - V^p_\varepsilon }}{{\sqrt {2{D(V^{p}_\varepsilon)}} }}} \right).
\end{aligned}
\end{equation}

\section{\label{sec5}Performance analysis}
The performances of both target detection and security check of the system are investigated in this section through numerical simulations based on the results from Secs. \ref{sec3} and \ref{sec4}.

In our simulations, we consider a realistic LiDAR system operating at a wavelength of 1550 nm with a shot-noise limited homodyne detector. The coherent detection in the LiDAR system allows it to inherently resist background noise, as the LO acts as a temporal, spatial, and spectral filter \cite{day}. Similar to a typical continuous variable quantum key distribution system through a free space channel, the excess noise and detector noise of the LiDAR system are relatively small compared to the shot noise, as characterized in the literature \cite{noise1}. In the following simulations, we set the reflectivity of the target to be $\eta_r=0.1$, and the detection efficiency and detection noise of the homodyne detector to be $\eta_d=1$ and $V_d=0.05$ SNU, respectively. Furthermore, we assume that the total excess noise of the system is less than 0.05 SNU.

\begin{figure}[!ht]
\centering
\includegraphics[width=1\columnwidth]{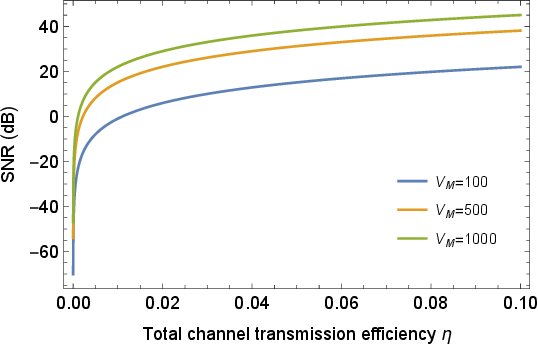}
\caption{The SNR versus total channel transmission efficiency. The curves (from bottom to top) are the results of signal modulation variance $V_M=100,500,1000$, respectively.}
\label{snrT}
\end{figure}

\begin{figure}[!ht]
\centering
\includegraphics[width=1\columnwidth]{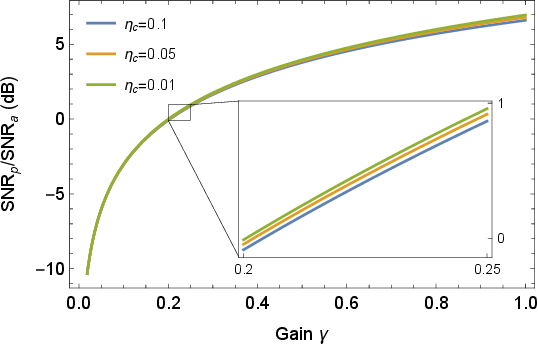}
\caption{The ratio of $\rm{SNR}_p$ to $\rm{SNR}_a$ as a function of the gain $\gamma$. The curves (from bottom to top) are the results of $\eta_c=0.1,0.05,0.01$, respectively.}
\label{Rsnr}
\end{figure}

Firstly, we investigate the SNR of the receiver in both cases. The ratio of $\rm{SNR}_p$ to $\rm{SNR}_a$ under different value of $\gamma$ is depicted in Fig. \ref{Rsnr} for channel transmissivity $\eta_c=0.1,0.05,0.01$, respectively.

In Fig. \ref{Rsnr}, we observe an increasing ratio of $\rm{SNR}_p$ to $\rm{SNR}_a$ as $\gamma$ increases. The differences in $\eta_c$ have only a small impact on the values of these ratios. Additionally, it's obvious that spoofing attacks decrease the SNR of the system when $\gamma<0.2$, but increase it when $\gamma>0.2$. Notably, the SNR remains unchanged when $\gamma=0.2$, indicating that the spoofer intentionally attenuates the measured data to compensate for the target's reflectivity and the 3 dB loss from heterodyne detection.

\subsection{Simulation results for target detection}

For target detection, we set the threshold of the correlation value to be $C_{th}=2\sqrt{V^{nf}_C}$. As a result, the false alarm rate is obtained to be $P_\text{fa}^T=0.023$. The probability of detection $P^T_\text{d}$ depends on the SNR, the data length L for target detection, and the phase drift parameter $\delta$.

\begin{figure}[!ht]
\centering
\includegraphics[width=1\columnwidth]{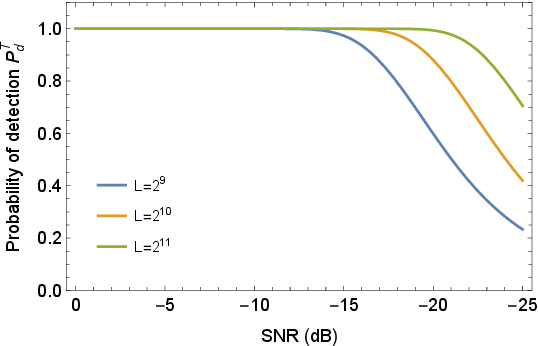}
\caption{The probability of detection $P_\text{d}^T$ for target detection as a function of SNR. Phase drift is set to 0. The curves (from bottom to top) are the results of $L=2^9,2^{10},2^{11}$, respectively.}
\label{PD-SNR}
\end{figure}

The probability of detection $P^T_\text{d}$ under different SNR conditions is depicted in Fig. \ref{PD-SNR} for $L=2^9,2^{10},2^{11}$, respectively, assuming $\delta=0$. As can be seen, the probability of detection decreases as the SNR decreases. However, increasing L can lead to a higher probability of detection by reducing the statistical fluctuation of the correlation value. When $L=2^{11}$, the probability of detection approaches 1 for SNR values greater than -20 dB. 

\begin{figure}[!ht]
\centering
\includegraphics[width=1\columnwidth]{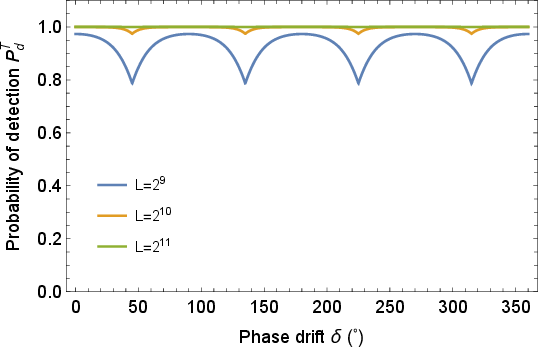}
\caption{The probability of detection $P_\text{d}^T$ for target detection versus phase drift with $\text{SNR}=-15$ dB. The curves (from bottom to top) are the results of $L=2^9,2^{10},2^{11}$, respectively.}
\label{PD-PS}
\end{figure}

In Fig. \ref{PD-PS}, the impact of different phase drifts $\delta$ on the probability of detection is demonstrated for SNR$=-15$ dB. It is obvious that the probability of detection reaches its lowest value when the phase drift is $\delta=45^\circ,135^\circ,225^\circ,315^\circ$, regardless the length $L$ for target detection. Moreover, the probability of detection remains unity for any phase drift with $L=2^{11}$.

\begin{figure}[!ht]
\centering
\includegraphics[width=1\columnwidth]{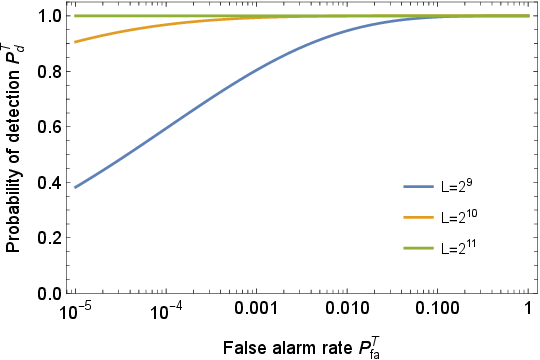}
\caption{ROC curves for target detection under different length of $L$ with $\delta=0$ and $\text{SNR}=-15$ dB. The curves (from bottom to top) are the results of $L=2^9,2^{10},2^{11}$, respectively.}
\label{TROCF}
\end{figure}

We also plot the receiver operating characteristic (ROC) curves in Fig. \ref{TROCF} based on Eq. (\ref{TROC}). It is apparent that a larger $L$ yields better performance. For SNR= -15 dB, $L=2^{11}$ is sufficient to detect the target with a very high probability. 

\subsection{Simulation results for security check}

The probability of detection ($P_\text{d}^S$) and false alarm rate ($P_\text{fa}^S$) are key metrics to characterize the security check operation of the LiDAR, using a predetermined threshold $\xi_{th}$.The parametric dependence of $P_\text{d}^S$ and $P_\text{fa}^S$ can be represented by the ROC curve. By selecting an appropriate excess noise threshold $\xi_{th}$, the security check operation of the LiDAR can be fine-tuned to achieve the desired probability of detection or false-alarm rate.

\begin{figure}[!ht]
\centering
\includegraphics[width=1\columnwidth]{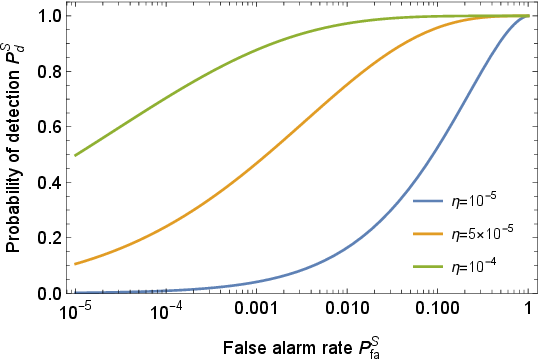}
\caption{ROC curves for security check under different total channel transmission efficiency $\eta$ with $M=10^6$ and $\gamma=0.2$. The curves (from bottom to top) are the results of $\eta=10^{-5},5\times10^{-5},10^{-4}$, respectively.}
\label{ROC1}
\end{figure}

Assume the number of the total interrogating pluses is $M=10^6$, and the spoofer chooses an gain of $\gamma=0.2$. Fig. \ref{ROC1} illustrates the ROC curves for channel transmission efficiency $\eta=10^{-5},5\times10^{-5},10^{-4}$. It is evident that a higher channel transmission efficiency leads to improved performance of the LiDAR system. For instance, when the probability of false alarm is set to $10^{-3}$, the probability of detection increases from less than 0.05 for $\eta=10^{-5}$ to approximately 0.91 for $\eta=10^{-4}$.

\begin{figure}[!ht]
\centering
\includegraphics[width=1\columnwidth]{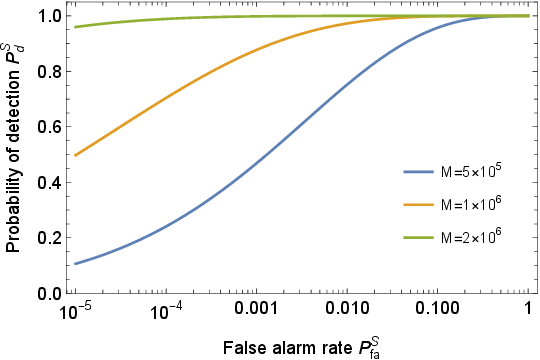}
\caption{ROC curves for security check under different interrogating signals $M$ with $\eta=10^{-4}$ and $\gamma=0.2$. The curves (from bottom to top) are the results of $M=5\times10^5,1\times10^6,2\times10^6$, respectively.}
\label{ROC2}
\end{figure}

Fig. \ref{ROC2} plots the the ROC curves with $M=5\times10^5,1\times10^6,2\times10^6$ when $\eta=10^{-4}$ and $\gamma=0.2$. It is clearly that the probability of detection increases with the number of interrogating pulses. When the number of pulses is increased to $2\times10^6$, the the probability of detection remains almost constant with a low value of false alarm probability, for example, $P^S_\text{fa}=10^{-4}$. This suggests that the spoofing attack can be detected reliably by the LiDAR system using $2\times10^6$ pulses to interrogate the target, with a low false alarm probability.

\begin{figure}[!ht]
\centering
\includegraphics[width=1\columnwidth]{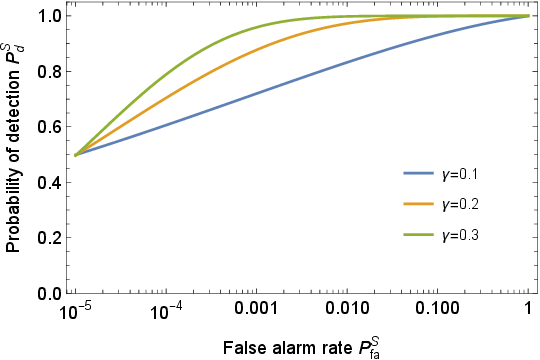}
\caption{ROC curves for security check under different gain $\gamma$ with $M=10^6$ and $\eta=10^{-4}$. The curves (from bottom to top) are the results of $\gamma=0.1,0.2,0.3$, respectively.}
\label{ROC3}
\end{figure}

\begin{figure}[!ht]
\centering
\includegraphics[width=1\columnwidth]{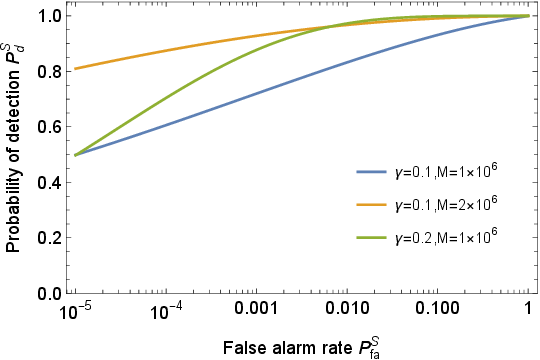}
\caption{ROC curves for security check under different gain $\gamma$ and number of interrogating signals $M$ with $\eta=10^{-4}$.}
\label{ROC4}
\end{figure}

Then we investigate the influence of the gain of the prepared signals induced by the spoofer on the ROC of the LiDAR system. Fig. \ref{ROC3} illustrates the ROC curves with $\gamma=0.1,0.2,0.3$ when $M=10^6$ and $\eta=10^{-4}$. The spoofer can degrade the probability of detection of the LiDAR system by attenuating the amplitudes of the prepared signals, due to the fact that the induced extra excess noise attenuated simultaneously with the signals, making it harder to be detected. Fortunately, this can be improved by increasing the number of interrogating pulses. As demonstrated in Fig. \ref{ROC4}, when the number of pulses is doubled, the probability of detection with $\gamma=0.1$ is dramatically improved, and it is even higher than the case with $\gamma=0.2$.

\section{\label{sec6}Conclusion}
To conclude, we have proposed a protocol for a quantum secured LiDAR system using Gaussian modulated coherent states. We have provided detailed models for both target detection and security check. The range can be determined by cross-correlating the signals, taking advantage of their inherent Gaussian nature. With a sufficient number of signals, such as $2^{11}$, the probability of detection for ranging can approach 1 even at low SNR, and the system is also resistant to phase drift. More importantly, we have shown that the LiDAR system can detect the IR spoofing attack performed by estimating the excess noise of the system. We have conducted numerical studies on the ROC of the LiDAR system for security check, considering different channel transmission efficiencies and total numbers of signals, as well as the gain of the prepared signals induced by the spoofer. Based on these results, when the channel transmission efficiency is $10^{-4}$ and the number of signals is $2\times10^6$, the LiDAR system can detect an IR spoofing attack with a probability of more than 0.999 at a low false-alarm rate of 0.001. This means that the integration time required for security check is only 10 ms for a LiDAR system with a repetition rate of 200 MHz. The spoofer's attenuation of the received signals may degrade the probability of detection. However, increasing the number of interrogating signals can effectively mitigate this issue. Moreover, the proposed protocol can be implemented using currently available technology, indicating its feasibility and practicality in real-world applications.

\appendix

\section{\label{app1} Covariance of $T_j$ and $R$}
Taking the phase drift into account, $T_1$ can be rewritten as 
\begin{equation}
\begin{aligned}
{T'_1} =& \left( {1 - B} \right) \cdot {X'_T} + B \cdot {P'_T}  \\
=& \left( {1 - B} \right)\left( {{X_T}\cos \delta  - {P_T}\sin \delta  } \right)\\
 &+ B\left({{P_T}\cos \delta   + {X_T}\sin \delta  } \right)\\
 =& {T_1}\cos\delta   + {T_2}\sin\delta  ,
\end{aligned}
\end{equation}
where $X'_T=A\cos\left(\theta+\delta \right)$ and $P'_T=A\sin\left(\theta+\delta \right)$. Then the result measured by the receiver is given by
\begin{equation}
R={\sqrt{\eta}}T'_1+N,
\end{equation}
where $N$ is the aggregated noise with variance $V_N$.
 
The covariance of $T_\nu$ ($\nu=1,2$) and $R$ is
\begin{equation}
\begin{aligned}
C_{{T_\nu}R} &= \left\langle {{T_\nu}R} \right\rangle  - \left\langle {{T_\nu}} \right\rangle \left\langle {R} \right\rangle =\left\langle {{T_\nu}R} \right\rangle \\
 &= \left\langle {{T_\nu}{\sqrt{\eta}}\left( {{T_1}\cos\delta  + {T_2}\sin\delta } \right) + {T_\nu}N} \right\rangle \\
& = {\sqrt{\eta}}\left\langle {T_\nu T_1} \right\rangle \cos\delta  + {\sqrt{\eta}}\left\langle {{T_\nu}{T_2}} \right\rangle \sin \delta .
\end{aligned}
\end{equation}
Considering that $\left\langle {T_1^2} \right\rangle =\left\langle {T_2^2} \right\rangle = {V_M}$ and $\left\langle {{T_1}{T_2}} \right\rangle =0$, we can obtain
\begin{equation}
\begin{aligned}
C_{{T_1}R} &={\sqrt{\eta}}V_M\cos\delta,\\
C_{{T_2}R} &={\sqrt{\eta}}V_M\sin\delta .
\end{aligned}
\end{equation}

\section{\label{app2} Noise floor of the cross-correlation profile}

The mismatch of $T_\nu$ and $R'$ results in an independent and identically distributed (i.i.d) noise floor, the cross-correlation of which is
\begin{equation}
 {{C^{nf}_{{T_\nu}R'}}}=\left\langle {{T_\nu}R'} \right\rangle =\left\langle {T_\nu} \right\rangle \left\langle R' \right\rangle=0. 
\end{equation}

With a finite length of $L$, $C_{T_\nu R'}$ can be estimated by the following unbiased estimator 
\begin{equation}
{\widehat {C^{nf}_{T_\nu R'}}}={\frac{1}{L}\sum\limits_{i = 1}^L {{T_{\nu i}}R{'_i}} }, 
\end{equation}
since
\begin{equation}
\left\langle  {\widehat {C^{nf}_{T_\nu R'}}} \right\rangle  = \frac{1}{L}\sum\limits_{i = 1}^L {\left\langle {{T_{\nu i}}{R'_i}} \right\rangle }  = \left\langle {{T_\nu}R'} \right\rangle  =  {{C^{nf}_{{T_\nu}R'}}}.
\end{equation}

For the covariance, we have
\begin{equation}
\begin{aligned}
V^{nf}_C & =D\left( {\widehat {C_{{T_\nu}R'}^{nf}}} \right) = \frac{1}{{{L^2}}}\sum\limits_{i = 1}^L {D\left( {{T_{\nu i}}R{'_i}} \right)}  = \frac{1}{L}D\left( {{T_\nu}R'} \right)\\
& = \frac{1}{L}\left( {\left\langle {T_\nu^2R{'^2}} \right\rangle  - {{\left\langle {{T_\nu}R'} \right\rangle }^2}} \right)\\
& = \frac{1}{L}\left( {\left\langle {T_\nu^2} \right\rangle \left\langle {R{'^2}} \right\rangle  - {{\left\langle {{T_\nu}} \right\rangle }^2}{{\left\langle {R'} \right\rangle }^2}} \right)\\
& = \frac{1}{L}{V_M}{({\eta}V_M+V_N)} \\
&=\frac{{{\eta}V_M^2}}{L}\left( {1 + \frac{{{V_N}}}{{{\eta}{V_M}}}} \right).
\end{aligned}
\end{equation}

\section{\label{app3} Peak value of the cross-correlation profile}

Similarly, the estimator of the cross-correlation peak value ${\widehat {C^p_{T_\nu R}}}$ is an unbiased estimator of $C^p_{T_\nu R}$, since
\begin{equation}
\begin{aligned}
\left\langle  {\widehat {C^p_{T_\nu R}}} \right\rangle  &= \frac{1}{L}\sum\limits_{i = 1}^L {\left\langle {{T_{\nu i}}{R_i}} \right\rangle }  = \left\langle {{T_\nu}R} \right\rangle  =  {{C^p_{{T_\nu}R}}}.
\end{aligned}
\end{equation}

For the covariance, we have
\begin{equation}
\begin{aligned}
V^p_{C\nu}&=D\left( {\widehat {C_{{T_\nu}R}^p}} \right) \\
=& \frac{1}{{{L^2}}}\sum\limits_{i = 1}^L {D\left( {{T_{\nu i}}{R_i}} \right)}  
= \frac{1}{L}D\left( {{T_\nu}R} \right) \\
=& \frac{1}{L}D\left[ {{T_\nu}{\sqrt{\eta}}\left( {{T_1}\cos \delta  + {T_2}\sin \delta } \right) + {T_\nu}N} \right]\\
=& \frac{{{\eta}}}{L}\left[ {D\left( {{T_\nu}{T_1}} \right)\cos^2 \delta  + D\left( {{T_\nu}{T_2}} \right)\sin^2 \delta } \right] + \frac{1}{L}D\left( {{T_\nu}N} \right) \\
=& \frac{{{\eta}}}{L}\left[ {\mu _\nu^2D\left( {T_\nu^2} \right) + \left( {1 - \mu _\nu^2} \right)D\left( {{T_1}{T_2}} \right) } \right] + \frac{{{V_M}{V_N}}}{{{L}}} \\
=& \frac{{{\eta}V_M^2}}{L}\left( { \mu _\nu^2 + 1 + \frac{{{V_N}}}{{{\eta}{V_M}}}} \right)\\
=& \frac{{{\eta}V_M^2}}{L} \mu^2_\nu +V^{nf}_C.
\end{aligned}
\end{equation}

\end{document}